\documentclass[11pt,twoside]{article}
\usepackage{asp2010}

\aspvolume{436}
 \aspcpryear{2010}
\aspvoltitle{Learning from Inquiry in Practice} 
 \aspvolauthor{L. Hunter and A. Metevier, eds.}

\resetcounters

\begin{document}

\title{The Adaptive Optics Summer School Laboratory Activities}
\author{S. Mark Ammons$^1$, Scott Severson$^2$, J.\ D.\ Armstrong$^3$, Ian Crossfield$^4$, Tuan Do$^4$, Mike Fitzgerald$^5$, David Harrington$^6$, Adam Hickenbotham$^7$, Jennifer Hunter$^8$, Jess Johnson$^9$, Luke Johnson$^9$, Kaccie Li$^7$, Jessica Lu$^{10}$, Holly Maness$^{7}$, Katie Morzinski$^9$, Andrew Norton$^9$, Nicole Putnam$^7$, Austin Roorda$^7$, Ethan Rossi$^7$, Sylvana Yelda$^4$
\affil{$^1$Hubble Fellow, Steward Observatory, 933 Cherry Ave., Tucson, AZ 85721}
\affil{$^2$Dept.\ of Physics \& Astronomy, Sonoma State University, 300L Darwin Hall, Rohnert Park, CA 94928}
\affil{$^3$Institute for Astronomy, University of Hawai`i, 34 Ohia Ku, Pukalani, HI 96768}
\affil{$^4$Department of Physics and Astronomy, University of California, Los Angeles, 430 Portola Plaza, Box 951547, Los Angeles, CA 90095}
\affil{$^5$Institute of Geophysics and Planetary Physics, Lawrence Livermore National Laboratory, L-413, 7000 East Avenue, Livermore, CA 94550}
\affil{$^6$Institute for Astronomy, University of Hawai`i, 2680 Woodlawn Drive, Honolulu, HI 96822}
\affil{$^7$University of California, Berkeley, Berkeley, CA}
\affil{$^8$Center for Visual Science, University of Rochester, Rochester, NY}
\affil{$^9$Center for Adaptive Optics, University of California, Santa Cruz, 1156 High St., Santa Cruz, CA 95064}
\affil{$^{10}$Department of Astronomy, California Institute of Technology, 1200 E.\ California Blvd., Pasadena, CA 91125}}

\begin{abstract} Adaptive Optics (AO) is a new and rapidly expanding field of instrumentation, yet astronomers, vision scientists, and general AO practitioners are largely unfamiliar with the root technologies crucial to AO systems.  The AO Summer School (AOSS), sponsored by the Center for Adaptive Optics, is a week-long course for training graduate students and postdoctoral researchers in the underlying theory, design, and use of AO systems.  AOSS participants include astronomers who expect to utilize AO data, vision scientists who will use AO instruments to conduct research, opticians and engineers who design AO systems, and users of high-bandwidth laser communication systems.  

In this article we describe new AOSS laboratory sessions implemented in 2006-2009 for nearly 250 students.  The activity goals include boosting familiarity with AO technologies, reinforcing knowledge of optical alignment techniques and the design of optical systems, and encouraging inquiry into critical scientific questions in vision science using AO systems as a research tool.   The activities are divided into three stations:  Vision Science, Fourier Optics, and the AO Demonstrator.  We briefly overview these activities, which are described fully in other articles in these conference proceedings (\cite{putnam_2010, do_2010, harrington_2010}, respectively).

We devote attention to the unique challenges encountered in the design of these activities, including the marriage of inquiry-like investigation techniques with complex content and the need to tune depth to a graduate- and PhD-level audience.  According to before-after surveys conducted in 2008, the vast majority of participants found that all activities were valuable to their careers, although direct experience with integrated, functional AO systems was particularly beneficial.
\end{abstract}

\section{Introduction:  Adaptive Optics Education}
Adaptive Optics is the technique of changing the wavefront properties of an optical system with a deformable mirror in response to measurements of wavefront quality.  This field of study is multi-disciplinary:  The techniques involved in modeling and constructing AO systems draw from engineering physics, control theory, optics, laser physics, and material sciences.  The systems themselves are relevant to the astronomical and biological sciences as well as to laser communications.  The diversity of experience of AO technicians and practitioners demands great care when educating the community on the principles of AO.  

This paper summarizes new laboratory sessions added to the Adaptive Optics Summer School in 2006-2009, tested on nearly 250 students.  Section 2 describes the larger Summer School and its lecture topics.  Section 3 discusses the original form of the lab activities in 2006.  Section 4 summarizes the final forms of the three activities used in 2007-2009, with emphasis on logistical details for the benefit of future laboratory coordinators.  Section 5 summarizes learner feedback gathered in 2008 and Section 6 discusses future improvements to make the labs self-sustaining.

\section{The Adaptive Optics Summer School}

The Center for Adaptive Optics (CfAO) is currently a University of California Multi-campus Research Unit, and was originally funded by the National Science Foundation (NSF) Science and Technology Center program from 1999-2009.  The CfAO Summer School program has been a significant component of the Center's NSF-mandated public outreach branch since 2000.  The goals of the summer school are to (1) broadly educate graduate students, postdoctoral researchers, and industry members about the theory, technique, and practical application of Adaptive Optics technologies and (2) to foster the growth of a large community of practitioners of AO in the parent fields of optics, astronomy, vision science, and laser communications.

The summer school is hosted annually in August on the campus of the University of California, Santa Cruz (UCSC).  The five day course is a compilation of four days of traditional lectures and 1.5 days of hands-on laboratory activities.  The size of the audience is capped at sixty graduate students, postdoctoral researchers, and industry members.

Recognized experts in the field of AO deliver twelve 1.5 hour lectures over the week.  The lectures range in interactivity from traditional talks with sparse interruption for questions to open-format panel discussions.  Lecturers are chosen by a committee of directors, typically themselves practitioners in AO research.  Viewgraph presentations are peer-reviewed by the lecturer pool.  Lecture topics include:
\begin{itemize}
\checklistitemize
\item Introduction to Adaptive Optics
\item The Principles of Wave Optics 
\item Adaptive Optics Control Theory and Practical Application
\item Adaptive Optics System Design
\item Wavefront Sensing and Reconstruction Techniques
\item Deformable Mirror Technologies and Implementation
\item Instrumentation for Adaptive Optics Systems
\item Adaptive Optics System Simulation
\item Applications of AO to Biology, Astronomy, and Laser Communications
\end{itemize}

\section{Laboratory Activity Development}
The AO Summer School program was augmented in 2006 with a series of laboratory activities developed through the CfAO Professional Development Program series of workshops (\cite{hunter_2010}, this volume).

\subsection{The Need for Laboratory Practice}
Students at AO Summer Schools prior to 2006 frequently commented that the majority of the lectures discussed physical phenomena as opposed to theoretical analysis, i.e., the interactions between deformable mirrors, wavefront sensors, and other AO system components.  A full understanding of an AO system requires an intimate knowledge of the nature of these interactions.  We began with the informed hypothesis that inquiry-like teaching techniques combined with direct experience with integrated, functional AO systems would benefit not only builders of AO systems, but their users as well.  

\subsection{A Diverse Audience}
The 60 students who attend the AO Summer School in a given year tend to be an equal mix of AO system builders and users.  Nearly two-thirds are graduate students and postdocs from CfAO partner institutions, with the majority of these intending to use existing AO systems in astronomy and vision science as tools to obtain scientific data.  The minority are tasked with maintaining or constructing new AO systems in their research.  One-third of the student pool are industry partners from observatories, national laboratories, and companies with interest in building and operating AO systems or conducting pure AO research.  A few faculty members attend who typically intend to begin programs utilizing AO instruments.

The student pool is widely diverse in age and experience level.  Many students have never worked with optics or lenses, while others are optics professors or researchers with decades of laboratory experience.  In designing the AO Summer School laboratory activities in 2006, we adopted two key approaches to address this range of learner backgrounds:
\begin{itemize}
\checklistitemize
\item \emph{Diverse Content Areas in Separate Activities.}  We chose to concentrate related topics into separate activities that cover a wide range of content altogether.  The three activities were:  The AO Demonstrator, AO Variables, and the Fourier Optics Computer Lab (descriptions below).  
\item \emph{Tiered Content Goals in Individual Activities.}  For individual activities, we developed a variety of content goals with a range of depths and difficulty levels, which were organized into tiers.  Inexperienced learners were expected to concentrate on the Tier 0 or 1 goals.  More experienced learners were directed to investigate the Tier 2 goals if it became clear through formative assessment that they had mastered Tier 0 and 1 goals. 
\end{itemize}
\subsection{Initial Goals}
With these guiding principles in mind, we developed the following content and scientific process goals for the activity series.  The overall theme of the instruction is the construction, optimization, and operation of AO systems.
\subsubsection{Content Goals}
\begin{enumerate}
\item \emph{Tier 0: Components of an Adaptive Optics System.}  Students are to recognize and understand the purpose of the deformable mirror, wavefront sensor, re-imaging optics, science camera, and control computer.
\item \emph{Tier 1: Manipulation of AO Variables.}  Students can explain the interactions between correction rate, gain, bandwidth, and final correction quality. 
\item \emph{Tier 1: Basic Optics Alignment.}  Learners display basic proficiency in optical alignment of a re-imaging system with lenses.
\item \emph{Tier 2: Interplay of AO System Components.}  Students can explain the need for a precise deformable mirror / lenslet array mapping and grasp the consequences of an imperfect mapping.  Students can determine the correct order of a set of re-imaging lenses in an AO system to achieve a good mapping.   Students understand the need for plane conjugation.
\item \emph{Tier 2: Relationship between Pupil Plane and Focal Plane.}  Learners can quickly guess the qualitative shape of a Point Spread Function (PSF) produced by a given pupil shape.  Learners can explain the effect of apodized pupils on the PSF.
\end{enumerate}

\subsubsection{Scientific Process Goals}
\begin{enumerate}
\checklistitemize
\item \emph{Tier 0: Confidence with Optics.}  Participants display simple confidence in the manipulation of optical components.
\item \emph{Tier 1: Operation of an AO System.}  Students can repeatably operate the controls of a working AO system to achieve basic functionality.  
\end{enumerate}
\subsection{Initial Implementation of Laboratory Activities}
In 2006, the summer school laboratory activities were divided into three stations:  The AO Demonstrator, AO Variables, and the Fourier Optics computer lab.  These activities were conducted at the Laboratory for Adaptive Optics (LAO) at UCSC.  

\subsubsection{AO Demonstrator}  The AO Demonstrator is a self-enclosed, standalone closed-loop adaptive optics system running at 30 Hz with an AgilOptics mirror (see Figure 1).  It was constructed at UCSC entirely for educational and demonstrative purposes.  It is composed of a 37-actuator drumhead deformable mirror, 8x8 Shack-Hartmann wavefront sensor, input laser, two sets of reimaging optics, and a science camera.  This camera, which records the focused laser beam as well as an inserted science image, sends output in real-time to a nearby screen.  A lens can be inserted into the beam to simulate an optical distortion.  When the Demonstrator is running in closed loop, the insertion of the distortion is corrected quickly and the science image returns to focus.  GUI software computes the deformable mirror commands from wavefront sensor signals.  

Groups of 3-4 students were shown the image quality achieved by the operating Demonstrator, then instructed to optically realign the Demonstrator after re-imaging lenses were removed.  Following realignment, their task was to achieve good closed-loop performance after automatic system calibration.  This activity is summarized in more detail in Harrington et al.\ (these proceedings).

The ease of realigning the AO Demonstrator ensured that the majority of groups succeeded in this task.  More advanced groups investigated the correction quality of the realigned system by changing control loop parameters or inserting new types of distortion into the optical beam (e.g., eyeglasses, plastic sheets).  This activity was retained and largely unchanged for later years.  

\subsubsection{AO Variables}  The AO Variables workbench was also a closed-loop AO system utilizing the Multi-Conjugate/Multi-Object AO research facility at the LAO.  This astronomical AO system included a high-order deformable mirror, a high-order Shack-Hartmann sensor, a science detector, a simulated Laser Guide Star point source, and a translatable atmospheric simulator.  Students were instructed not to interact with the optics in this AO system to prevent system failure.  In this activity, groups of 3-4 students directed the AO system to perform in closed loop with control over two variables:  Correction rate and gain.  Using data from the science detector and measured rejection transfer functions, students optimized the correction rate and gain for a given guide star magnitude and wind speed.  The participants were expected to discover that decreasing guide star brightness requires slower correction rates.

Unfortunately, due to complexities of the underlying AO system, this activity was less successful than others.  The correction quality of this system began to degrade over time, which was difficult to explain to participants during the activity.  In addition, students reported disappointment that they were not permitted to interact with the optics and system components.  This activity was not retained for later years.  

\subsubsection{Fourier Optics Computer Lab}  The Fourier Optics activity was first implemented as a computer lab, conducted in parallel in groups of two.  Students were given an IDL GUI that allowed the user to compute a modular transfer function (MTF) and point spread function (PSF) for a variety of pupil shapes, with and without the presence of atmospheric turbulence.  The groups were given worksheets with instructions on operating the GUI and were generally left to pursue interesting questions on their own.  This station was later modified to include optical components, as described in Section 4.2.

\section{Final Form of Laboratory Activities}
The laboratory activities were standardized from 2007-2009 into three stations:  Fourier Optics, AO Demonstrator, and Vision Science.  These activities are summarized below and are described in more detail in these proceedings (\cite{do_2010, harrington_2010, putnam_2010}).  

\subsection{Adaptive Optics Demonstrator}

The AO Demonstrator station was largely unchanged from 2006 (see Section 3.4.1).   In later years, more AO Demonstrators were constructed and used in parallel at the summer school.  Two Demonstrators of the AgilOptics type currently exist at UCSC and Maui Community College.  Two Demonstrators utilizing an Iris AO deformable mirror currently exist at Hawaii Community College and the Institute for Astronomy at the University of Hawaii.  These systems have been used at various times (frequently shipped across the Pacific) in the years 2007-2009.  The use of the AO Demonstrator in optics education is discussed in more detail in \cite{harrington_2010} . 
\begin{figure}
\begin{center}
\includegraphics[scale=0.55]{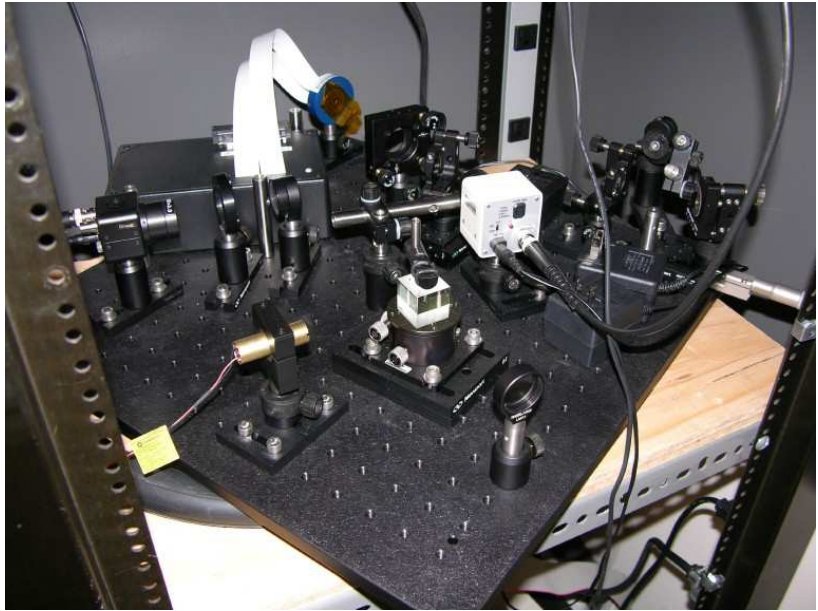}
\caption{The Adaptive Optics Demonstrator.  The blue covered component is the AgilOptics deformable mirror and the central white box is the science detector. }
\label{FO}
\end{center}
\end{figure}

\subsection{Fourier Optics}

The Fourier Optics computer lab was modified to include optical components and an improved GUI, written by T. Do.  The optical setup, shown in Figure 2, includes a 1 meter rail, a point source laser, two imaging optics, an iris or other pupil stop, and a detector.  A laptop is used to display the image from the detector in real time.  Participants work out relationships between features in the PSF and the shape of the pupil.  The pupil can be manipulated by cutting holes of various shapes in card stock.  Students are also given a small lenslet array and asked to construct a Shack-Hartmann wavefront sensor from the optics provided.  Following these activities, students are encouraged to investigate high-contrast imaging systems or other novel wavefront sensing systems covered in summer school lectures.  Students also have access to the Fourier Optics GUI, which computes PSFs for different pupil shapes (with and without turbulence).  Further details of the optical layout and the implementation of this activity are given in \cite{do_2010}.
\begin{figure}
\begin{center}
\includegraphics[scale=0.55]{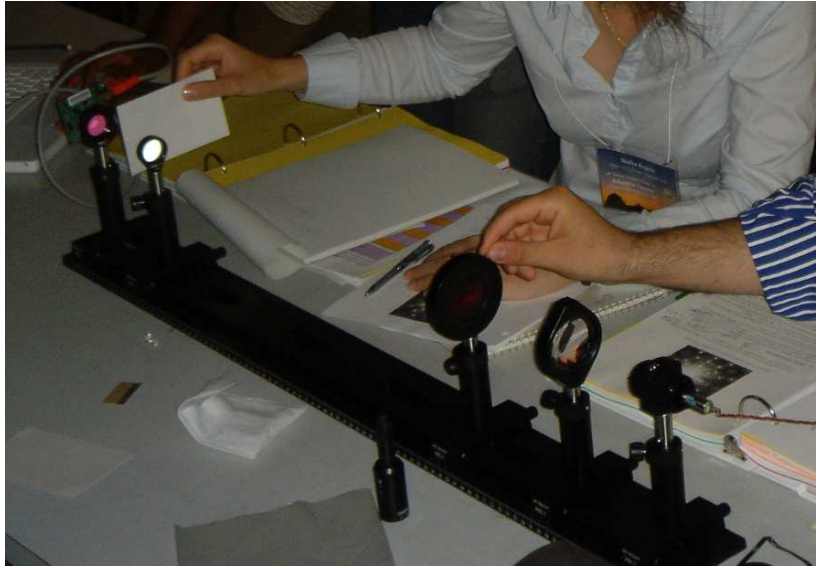}
\caption{The Fourier Optics apparatus in use in 2009.  The system is composed of a point source, several imaging lenses, an iris or pupil of adjustable shape, and an imaging detector mounted on a rail.}
\label{AOD}
\end{center}
\end{figure}

\subsection{Vision Science Activity}
In the Vision Science activity (see \cite{putnam_2010} in these proceedings for more), students investigate pre-written questions involving human vision using inquiry-like techniques.  Students use the wavefronts of their own eyes as data (measured concurrently) to construct informed answers.  Over three hours, groups of three students view ``starter'' demonstrations and answer two separate questions in two periods of focused investigation.  The questions addressed include:
\begin{enumerate}
\item How does pupil size affect vision? How does vision/environment affect pupil size? 
\item What is your depth of focus? How does refractive error and pupil size affect depth of focus?
\item What is accommodation? What is your amplitude of accommodation? Does accommodation change with age? 
\item How do you determine refractive error? What is your refractive error? What is it like to be nearsighted or farsighted? 
\item What is astigmatism and how does it affect vision? How is astigmatism corrected? 
\item Is there a relationship between the aberrations of my right \& left eyes? 
\item Is there a relationship between a Zernike termÕs root-mean-square and its radial order? 
\item How do different aberrations interact with one another? Do they add together or can they cancel each other out?
\end{enumerate}
An additional tool available to learners is the Fourier Optics GUI, which allows users to load wavefronts of the eye, compute the resulting PSFs, and convolve them with a scene.  At the end of the activity, participants create posters summarizing their question/investigation and share to the larger group (see Figure 3 for an example poster).  

\begin{figure}
\begin{center}
\includegraphics[scale=0.40]{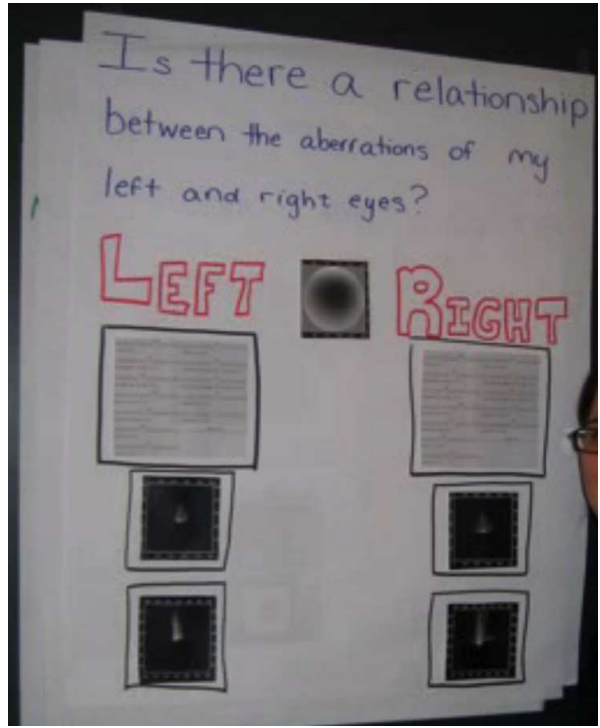}
\caption{Example poster completed by a summer school team in 2009.  In this example, students use computed point spread functions to show that left and right eye wavefronts are mirror images of each other. }
\label{VS}
\end{center}
\end{figure}

\begin{figure}[!htb]
\begin{center}
\includegraphics[scale=0.42]{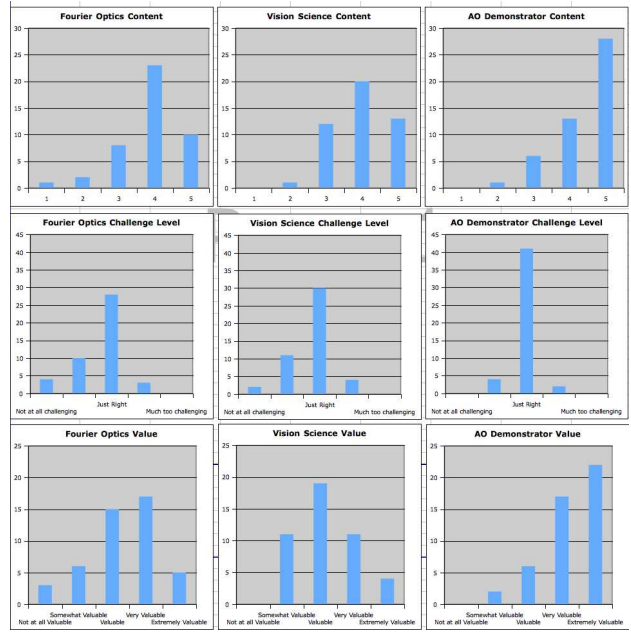}
\caption{Summary of selected student feedback responses from the 2008 Summer School.  Survey questions are arranged in rows and individual activities are arranged in columns, with FO corresponding to the Fourier Optics activity, VS corresponding to the Vision Science Activity, and AOD corresponding to the AO Demonstrator Activity.  }
\label{feedback}
\end{center}
\end{figure}

\section{Learner Feedback}

The AOSS laboratory team has primarily relied on learner feedback to inform changes from year to year.  For each station, students are asked to rate the content, quality of instruction, level of challenge, format, and overall value on a 1 to 5 scale.  The results on three of these metrics for 2008 are shown as arrays of bar graphs in Figure 4.   

For Fourier Optics, the mean content rating was 3.89 out of 5; 64\% of respondees rated the level of challenge as ``just right;'' and 50\% rated the activity as ``very valuable'' or higher.  For Vision Science, the mean content rating was 3.98 out of 5; 65\% of respondees rated the level of challenge as ``just right;'' and 32\% rated the activity as ``very valuable'' or higher.  For the AO Demonstrator, the mean content rating was 4.42 out of 5; 87\% of respondees rated the level of challenge as ``just right;'' and 83\% rated the activity as ``very valuable'' or higher.

All three of the activities are rated highly for content quality, level of challenge, and overall value.  The majority of respondents praise each activity as valuable to their future careers in written comments.  However, the numerical results shown above indicate that experience with the AO Demonstrator is particularly considered worthwhile.  We conclude that direct interaction with integrated, functional AO systems is notably beneficial to AO Summer School participants.

\section{The Summer School Laboratory Activities as a Self-Sustaining Entity}
Our principal vision for the summer school laboratory activities is that they become self-sustaining.  The achieve this goal, it is important to build a community of experienced facilitators (instructors) at the campuses of the University of California, which fund the CfAO and the summer school.  This will be encouraged by implementing honorariums for lab facilitators.  Other venues, e.g., graduate AO courses, are increasingly utilizing these activities as well, ensuring that specific facilitation techniques are not forgotten from year to year.  Secondly, it is important that the summer school independently own all necessary equipment, so that critical parts do not need to be shipped.  All equipment for Fourier Optics and Vision Science is located at either UCSC or UC Berkeley.  In 2010, two new AO Demonstrators will be constructed at UCSC and Iris AO/Berkeley.  

\acknowledgments
The authors wish to acknowledge considerable support for the laboratory additions by current and previous CfAO administrators and AO Summer School directors.  These include Julian Christou, Michael Fitzgerald, Donald Gavel, Chris Le Maistre, Karen Pe\~{n}a, Scott Severson, Paula Towle, and Leslie Ward.  This material is based upon work supported by: the National Science
Foundation (NSF) Science and Technology Center program through the Center for Adaptive Optics, managed by the University of California at Santa Cruz (UCSC) under cooperative agreement AST\#9876783; UCSC Institute for Scientist \& Engineer Educators.

\bibliography{ammons}
\bibliographystyle{asp2010}
\end{document}